\documentclass[a4paper,fleqn]{cas-dc}

\usepackage[square,sort,comma,numbers]{natbib}
\setcitestyle{square}
\usepackage{tabularx}
\usepackage{tikz}
\usepackage{caption}
\usepackage{multirow}
\usepackage{listings}
\usepackage{tikz,pgfplots,pgfplotstable}
\usepackage{xurl}
\usepackage{algorithm}
\usepackage{cryptocode}
\newcommand{\ballnumber}[1]{\tikz[baseline=(myanchor.base)] \node[circle,fill=.,inner sep=1pt] (myanchor) {\color{-.}\bfseries\footnotesize #1};}

\usepackage{amssymb}

\usepackage{pifont}
\usepackage{hyperref}

\newcommand{\cmark}{\ding{51}}%
\newcommand{\xmark}{\ding{55}}%

\begin{document}
\let\WriteBookmarks\relax
\def\floatpagepagefraction{1}
\def\textpagefraction{.001}
\shorttitle{Addressing Network Packet-based Cheats in Multiplayer Games: A Secret Sharing Approach}
\shortauthors{Y. Cai, K. Markantonakis, and C. Shepherd}

\title [mode = title]{Addressing Network Packet-based Cheats in Multiplayer Games: A Secret Sharing Approach}

\author[1]{Yaqi Cai}
\ead[url]{yaqicai@outlook.com}

\credit{Conceptualization of this study, Methodology, Software}


\author[1]{Konstantinos Markantonakis}
\ead[url]{k.markantonakis@rhul.ac.uk}

\author[2]{Carlton Shepherd}[orcid=0000-0002-7366-9034]
\cormark[1]
\ead[url]{carlton.shepherd@ncl.ac.uk}

\address[1]{Information Security Group, Royal Holloway, University of London, Egham, Surrey, United Kingdom}

\address[2]{School of Computing, Newcastle University, Newcastle-upon-Tyne, United Kingdom}

\cortext[cor1]{Corresponding author}

\tnotetext[1]{C.\ Shepherd has received funding from the UK EPSRC `Chameleon' project (EP/Y030168/1).}

\begin{abstract}
 Multiplayer online gaming has witnessed an explosion in popularity over the past two decades. However, security issues continue to give rise to in-game cheating, deterring honest gameplay, detracting from user experience, and ultimately bringing financial harm to game developers. In this paper, we present a new approach for detecting network packet-based cheats, such as forgery and timing cheats, within the context of multiplayer games using an application of secret sharing. Our developed protocols are subjected to formal verification using AVISPA, and we present simulation results using a Python-based implementation. We show that our proposal is practical in addressing some widely used attacks in online gaming.
\end{abstract}


\maketitle

\section{Introduction}

\par According to Statista \cite{Statista_2018}, the revenue of the digital video game market surpassed \$250 billion USD in 2023. Alongside the increasing popularity of online games, cheating is an ever-present, pernicious force that undermines the integrity and fairness of gameplay. According to the survey conducted by Irdeto, a cyber security services firm,  88\% of the interviewed players have had negative gaming experiences associated with cheaters \cite{granados_2018}. Dishonest players may use various cheating methods such as altering the game files, manipulating network packets, and in-memory object in order to disrupt the gameplay. Such activities negatively affects other honest players, leading to lower game traffic and a reduction in revenue, which has prompted research on online game cheating.
\par  Several theoretical frameworks have been proposed to prevent and address online game cheats with varying advantages and disadvantages~\cite{racs, pp-ca, lockstep} (see \S\ref{sec:relatedwork-discussion}). Commercial anti-cheat software, such as Easy Anti-Cheat \cite{eac} and BattlEye \cite{battleye}, are available, focussing principally on operating system-level controls, e.g.\ DLL hook detection and periodic stack checks, and using signature-based approaches for detecting malicious processes. We observe that commercial products are closed-source, largely black-box solution, making it difficult to understand their precise security goals and countermeasures. We also note that some commercial anti-cheat engines have also been shown to be vulnerable to attacks themselves (e.g.\ see \cite{guigo2014next}). 
\par In this paper, we evaluate and build upon the current state of the art in online cheating. We propose a new approach for cheat detection in online multiplayer gaming that unifies the distinct approaches of RACS proposal by Webb et al.~\cite{racs} and Matsumoto-Okabe~\cite{Matsumoto}, building a new scheme with protection against collusion, packet inconsistency attacks and timing cheats in online gaming scenarios.  

\label{scope}
This paper provides the following contributions: 
\begin{itemize}
\item An evaluation of state-of-the-art cheating methods in online multiplayer gaming.
\item A novel approach for detecting network cheats in multiplayer online games using secret sharing.
\item An evaluation and verification of the developed protocols using AVISPA~\cite{armando2005avispa} and simulation results using a Python implementation.
\end{itemize}



The remainder of this paper is described as follows. \S\ref{ch:background} presents a primer on cheating in online games and discusses cheats that the paper aims to address, and common multiplayer game architectures. \S\ref{ch:relatedwork} analyses existing literature and their strengths and weaknesses in addressing the threats from \S\ref{ch:background}. Next, \S\ref{ch:model} proposes an improved theoretical framework that builds upon the previous works covering an extensive range of game session areas. \S\ref{ch:evaluation} conducts a simulation of our proposed framework and evaluates its security performance using a protocol verification tool AVISPA. \S\ref{ch:simulation} describes the simulation in Python and shows the average execution time and the physical memory usage for each process. Lastly, we conclude this paper, summarise the main findings, and outline areas of further research.

\section{Background}
\label{ch:background}
We adopt the definition of cheating given by Yan and Randell~\cite{YanandRandell}. Namely, we define a \emph{cheater} as a player whose in-game behaviour satisfies two conditions:
\begin{enumerate}
    \item \textbf{Unfair}: Refers to actions or behaviors that provide an individual with an advantage over others in a manner that disrupts the intended balance or fairness of a system player gains an advantage over other players. These actions undermine the equitable experience designed by the system creators, even if they don not necessarily involve breaking into the system or violating explicit access controls.
    \item \textbf{Unauthorised}: The unfair behaviour from (1) violates the intended controls established by the game's developers. It may involve, for example, accessing, modifying, or using system resources without the necessary permissions or approval from the game's developers or operators.
\end{enumerate}

Note that only satisfying one condition does not indicate that the player is a cheater. For example, a player may purchase high-quality peripherals to gain an additional advantage. While this can create an unfair competing environment for those players who do not have such peripherals, it is not strictly prohibited by the game operators. Similarly, users who express discourtesy or explicit language towards other players. While this is strongly discouraged, and it can be considered unethical, it does not necessarily offer any advantages to the player(s), and it therefore cannot be considered as cheating.
\subsection{Cheats Classification}
\label{classification}
Several papers have aimed to classify and taxonomise online game cheats \cite{pritchard_2000,YanandRandell,Yan,racs,feijoo2012mobile}. In this paper, we adopt the classification proposed by Webb et al.~\cite{racs}, which classifies online cheats into four levels: \ballnumber{1} game-level, e.g.\ bug exploitation; \ballnumber{2} application cheats, e.g.\ auto-aiming and peripheral spoofing; \ballnumber{3} protocol-level, e.g.\ packet forgery; and \ballnumber{4} infrastructure, e.g.\ information leakage from databases.

In this work, we specifically aim to address cheats in \ballnumber{3} which involve observing/sniffing, intercepting, modifying, and delaying the network packets sent during the course of a game in an unauthorised fashion. Packet forgery, timing, and collusion are the three primary sub-types of such cheats. We discuss each of these in the following subsections.

\subsubsection{Packet Forgery}

In \emph{packet modification} cheats, a cheater may modify the network packets and claim that the results contained in the tampered packets are correct. Myeongjae~\cite{Myeongjae_2019} showed how a cheater may edit the relevant traffic packets and sends false information to the server.  In this work, cheaters may also aim to induce \emph{inconsistency} issues in which cheaters may send incorrect packets only to a specific honest player in the context of games that permit network communication between players. This can induce a state of inconsistency that corrupts the game state of the honest player, who will be subsequently disadvantaged in the game. Detecting such attacks can be difficult; honest players may be unable to detect if it is a malicious attack or an unintentional corruption~\cite{NEO}. 

To tackle this problem, a fundamental solution is to involve a trusted third party that brokers all communications and maintains a consistent game state for all players ~\cite{Corman}. Endo et al.~\cite{endo2006cheat} also introduce alternative mitigation called the group votes, which creates a consensus of a consistent game state using a majority voting system. However, Webb et al.~\cite{racs} raised concerns that cheaters may collude in calculating a majority vote, allowing a significant subset of players to gain an advantage.

\subsubsection{Timing Cheats}
\label{sec:timing-cheats}

Alternatively, cheaters may intentionally intercept their outgoing packets while still receiving incoming packets in order to gain additional time and information. These factors may allow the cheater to make a more optimal decision relative to other players. Specifically, a cheater can conduct the following timing-based cheats. Firstly, \emph{suppressed updates} involve cheating players that intentionally drop their outgoing network packets relevant to the game in question. Considering the fact that some players may have an unstable network connection, some online games use dead reckoning-based methods to simulate the positions of unresponsive players~\cite{Aronson_1997}. However, a cheater may abuse this technique by purposely dropping $n-1$ consecutive updates, where $n$ is the largest number of packets that dead-reckoning allows players to drop. Based on the received incoming updates, the cheater calculates his optimal decision and sends it at the \(n^{th}\) packet to avoid being disconnected by the server. Once the updates are simulated, cheaters may suddenly appear in another place that honest players will not expect. 

Secondly, cheaters may use \emph{fixed delays}, where they intentionally delay their outgoing packets. Nichols and Claypool~\cite{Nichols} have observed this cheating behavior in the Madden NFL Football video game, enabling players to gain a significant advantage. In addition to the unfair advantages gained by using suppressed updates, a fixed delay may introduce a high level of latency to all other players.  

\subsubsection{Collusion}

Collusion cheats more generally occur when two or more players collude to secretly exchange information. Suppose three players $\{P_1, P_2, P_3\} \subseteq \textbf{P}$ are in the same game session, where $\textbf{P}$ is the set of \emph{all} game players within that session. $P_1$ is able to communicate with $P_2$, while $P_3$ is blocked by an immutable object with some occluding properties. If $P_3$ is an honest player, then he cannot see the other two players (nor is he able to communicate with other players). However, if $P_2$ transmits his incoming packets simultaneously to $P_3$, then $P_3$ may learn additional information that he is not supposed to know (e.g.\ about the position of $P_1$ who is then placed at a disadvantage).



\subsection{Multi-Player Game Architecture}
\par The client-server (C/S) architecture is the most common, and arguably simplest, network architecture used in multiplayer online games~\cite{article}. An authoritative server exists in this environment while other parties are considered as untrusted clients, i.e.\ corresponding to the players' machines. The server holds global knowledge of the game world and is able to calculate and communicate the game state with clients. The clients are only responsible for processing the reply from the server and rendering the game world on individual players' displays. If a client's game state is changed, the server will update the latest information to the client and inform other players on behalf of him.


In comparison, a peer-to-peer (P2P) architecture is decentralised in nature. Rather than using a single trusted server to maintain the game state, each node in the network topology is considered as a peer with the same capabilities and responsibilities. Each peer controls its own game state; it is effectively able to act as a server or a client, depending on whether the player is requesting or replying to the message. The player constructs the game state individually which is communicated with other players in order to attain consensus about a common agreed game state.
\section{Related Work}
\label{ch:relatedwork}

This section discusses several proposed frameworks that have been for preventing multiplayer online cheating and their relative advantages and disadvantages.

\subsection{Lockstep}
\begin{figure}
	\centering
		\includegraphics[scale=.7]{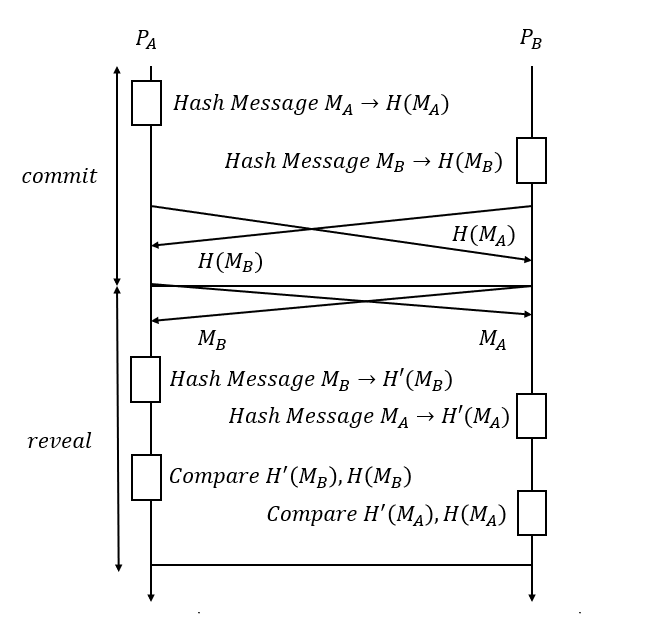}
	\caption{Lockstep protocol involving two players, $P_A$ and $P_B$.}
	\label{FIG:lockstep}
\end{figure}

Lockstep is a stop-and-wait protocol with a decision commitment step \cite{lockstep}, which addresses timing-based cheats where a cheater delays the announcement of his actions until all other players have sent their decisions. This type of cheat is also referred to as a \emph{look-ahead} cheat. An overview of this protocol is illustrated in Figure~\ref{FIG:lockstep}, comprising two phases: the \emph{commit} and \emph{reveal} phase. In the commit phase, player $P_i$ hashes his actions encoded in a message, $M_i$, using an appropriate cryptographic hash function, $H$, and sends $H(M_i\, ||\, s_i)$ to all other players in the game, where $s_i$ is an optional random salt value and $||$ is the concatenation function. Only after all players have announced their hashed actions will players enter the reveal phase. Then, $P_i$ sends the action $M_i$ (and optional salt, $s_i$) in plaintext and verifies other players' decisions by comparing the values he received in the previous phase. The Lockstep protocol addresses the look-ahead cheat as the hashed message prevents cheaters from gaining any additional information in the commit phase.  

However, one of its greatest drawbacks of Lockstep is that the design of the commit phase introduces a performance penalty. If there is a player with high latency, the game can become unacceptably slow as everyone must wait for his announcement before entering the reveal phase. As such, the protocol's performance is limited to a worst-case scenario by waiting for the player with the highest latency before entering the next step of the game.

\subsubsection{Asynchronous Synchronisation (AS) \label{AS}}

\begin{figure}[h]
	\centering
		\includegraphics[scale=.5]{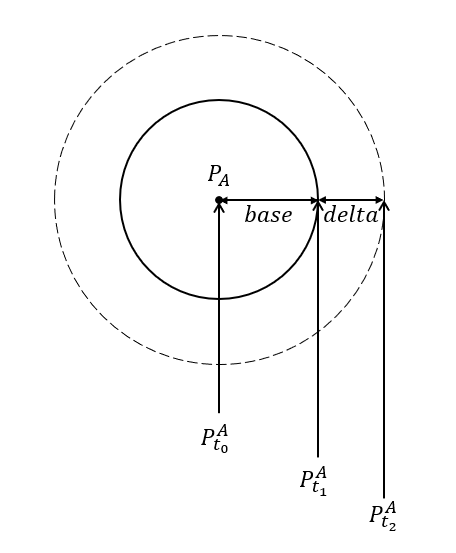}
	\caption{Spheres of influence in~\cite{AS}.}
	\label{FIG:AS}
\end{figure}

To mitigate the performance penalty of Lockstep, Baughman and Levine~\cite{AS} proposed the Asynchronous Synchronisation (AS) scheme. In this proposal, each player prepares his announcement asynchronously and only enters into a Lockstep-style scheme when someone else is within the player’s \textit{sphere of influence} (SoI). An SoI, shown in Figure~\ref{FIG:AS}, is an area in which the player's decisions can have influences on other players. Suppose a player $P_1$ is in an arbitrary turn of a game, $t_0$. The area of the circle in the solid line is the area that may be influenced by $P_1$'s announcements on the next turn $t_{1}$ and the area of the circle in the dashed line is the influence area in subsequent turns, e.g.\ $t_{2}$. The \textit{base} and \textit{delta} represent the radius of the circle in the solid line and the one in the dashed line respectively across turns.

\begin{figure}
    \centering
		\includegraphics[scale=.52]{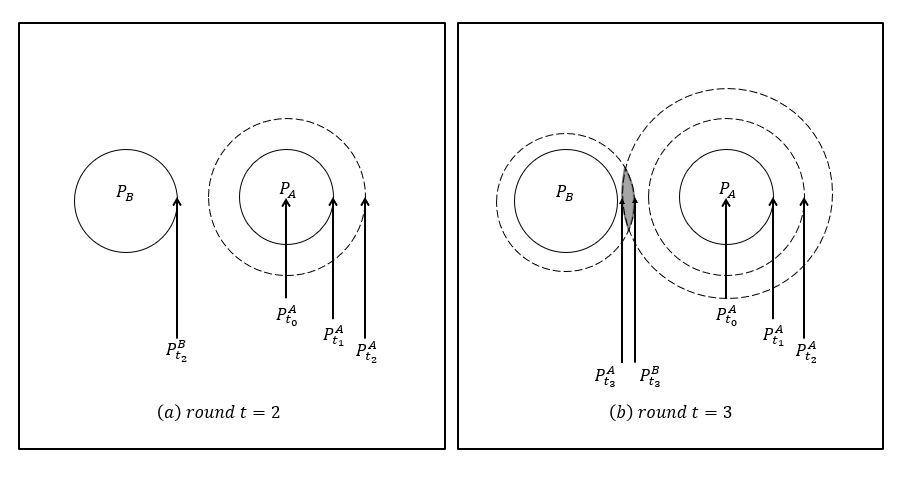}
	\caption{Spheres of Influences: Interaction.}
    \label{fig:AS-2}
\end{figure}

During the commitment phase of AS, a player $P_1$ commits his announcement to all the players in advance, and checks who is supposed to receive the message. If there is no intersection between two players' SoI, then $P_1$ will not interact with $P_2$ at some time step (e.g.\ $t=2$ in Figure~\ref{fig:AS-2}). However, if $P_2$ \emph{does} enter $P_1$'s area of interest in a subsequent time step, e.g.\ $t=3$ in Figure~\ref{fig:AS-2}, they will enter into a Lockstep model. Note that, while a player will no longer be required to interact with all other in-game players, the time for exchanging messages remains the same. Thus, if a player, $P_i$ has high latency, then the gameplay experience of the players in $P_i$'s SoI will be affected.

\subsection{Peer-to-Peer with Central Arbiter (PP-CA)}

Peer-to-Peer with Central Arbiter (PP-CA) is a protocol proposed by Pellegrino and Dovrolis~\cite{pp-ca} that uses a hybrid architecture between C/S and P2P and aims to provide high game state consistency with low bandwidth requirements. In this proposal, the authors introduce a new entity, \emph{the central arbiter}, within the traditional P2P architecture. The role of the central arbiter is to monitor the communication among players, simulate gameplay, and detect any inconsistencies. In contrast to a single trusted server (cf.\ C/S architectures), the central arbiter only sends a corrected update to players when it detects any inconsistencies. While PP-CA attempts to the features of C/S and peer-to-peer architectures, the authors only focused on its computational performance and ignored its cheats resistance. For instance, the definition of inconsistency remains unclear; it is not definitive as to what inconsistency events may trigger updates by the arbiter. It is conceivable that if a large number of game events are deemed inconsistent, then the benefits over a traditional C/S are significantly diminished. 
 
\subsubsection{Referee Anti-Cheat Scheme}
\label{RACS}
\begin{figure}
    \centering
        \includegraphics[scale=.65]{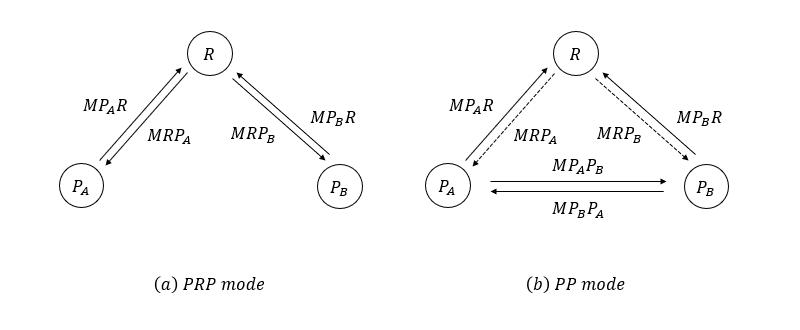}
    \caption{RACS communication models.}
    \label{fig:RACS}
\end{figure}

Webb et al.~\cite{racs} extended PP-CA in the Referee Anti-Cheat Scheme (RACS). Here, a trusted referee, $R$, is introduced that communicates updates to players and maintains the authorised game state. RACS comprises has two communication models, as shown in Figure~\ref{fig:RACS}. Firstly, Peer-Referee-Peer (PRP) mode is similar to C/S architecture: $R$ verifies and mediates network traffic between the peers, $P_A$ and $P_B$. $P_A$ updates and sends his game states $MP_AR$ to $R$. Next, $P_A$ receives \(MRP_A\) from $R$, which is the game state agreed with $R$ with respect to peers within an `area of interest' (AoI). Once the peers are updated, $R$ instructs them to enter Peer-Peer (PP) mode. In this mode, $P_A$ updates his game states $MP_AP_B$ to his peer $P_B$ directly, sending a copy of the update to $R$. The referee only contacts players once a player has disconnected or been detected as a cheater.

\par RACS is argued to provide cheat resistance equivalent to a C/S architecture with a lower bandwidth requirement; however, there are several challenges. Firstly, although the AoI model is implemented in this protocol, its mechanism and definition are not specified. Moreover, it does not protect against collusion attacks. Suppose there is a third player, $P_C$, who is not in the AoI of $P_A$ and $P_B$. If $P_B$ colludes with $P_C$, and sends $P_A$'s update, $MP_AP_B$, to $P_C$ secretly, then $P_C$ will receive additional, unauthorised information with which he may gain an unfair advantage. 
\subsection{Majority Votes}

Endo et al.~\cite{endo2006cheat} proposed a solution to the inconsistency problem using majority votes. Some clients are selected by the server as `super nodes' to manage and calculate the same game data. Once the super nodes have completed their calculations, they update their results to a trusted server, which determines the final state if it is agreed by at least half of such nodes. The clients involved in calculating the game state are chosen from distinct regions of the game such that, even even if the clients are malicious, they do not gain much benefit from tampering attempts. However, Matsumoto and Okabe~\cite{Matsumoto} questioned whether the fixed threshold value (i.e.\ half of the selected clients) is secure enough against a collusion attack. For instance, cheaters can use artificial clients (e.g.\ though a botnet) to increase the success rates of collusion attacks. Another drawback is cheaters may collude across regions: although super nodes are managing data in other regions, they may still contact players in those regions. 

\subsubsection{The Matsumoto-Okabe Framework}
\label{collusion}
Matsumoto and Okabe~\cite{Matsumoto} proposed an improved majority vote framework. Here, a player, $P_A$, sends his encrypted game state, $Enc(X)$, to a trusted server, $S$. The server decrypts the message and sends a share\footnote{By a `share' we refer to the outputs of a secret sharing scheme, e.g.\ a linear scheme; see \S\ref{sec:sss} and Shamir~\cite{shamir1979share}.}  of the game state sent by \(P_A\) to each related player including \(P_A\). This process is repeated for each player, who calculates his events and returns the results to the other players. The next game state is reconstructed from $k$ shares, where $k$ is a threshold determined by $S$. However, such a scheme may be exploited by sending incorrect results in the form of a colluding parties. For example, suppose that two players, $\{P_B,P_C\}$, are cheaters within a particular game in the vicinity of \(P_A\). The players may send malicious states to \(P_A\) via $S$. Because $S$ is a mediator and does not actually validate the game state, \(P_A\) may reconstruct a malicious game state from the shares sent by $\{P_B,P_C\}$.

\subsection{Discussion}
\label{sec:relatedwork-discussion}
In this work, we contribute new methods for preventing against packet forgery, timing, and collusion cheats.  In the following discussion,  \(P_A\) is assumed to be an honest player while \(P_B\) is a cheater. Table~\ref{tab:protocol} is presented to summarise our evaluation. 

 \emph{Packet forgery} cheats can be prevented by all evaluated protocols. In AS, an honest player can identify this cheating by comparing the hash value he received (\(H(M_B)\)) and the hash of the received plaintext (\(H'(M_B)\)). If a cheater sends a modified packet, these two values will be inconsistent. RACS has a similar solution in which \(R\) compares the hashes of all the messages received by \(P_A\) in the previous round with those received by \(R\) from \(P_B\). We note that, as RACS is limited more generally to providing the same security features as C/S architecture. \emph{Inconsistency cheats} can be addressed by all protocols except AS. In AS, the lack of a trusted entity indicates that there is no way to validate universally whether inconsistent packets are intentionally sent by malicious players. In comparison, RACS can prevent it due to the existence of a trusted referee \(R\). Inconsistency attacks does not exist in CRT because the player will not send the same packets to different players.

Recall from \S\ref{sec:timing-cheats} that timing cheats may involve \emph{suppressed updates} and \emph{intentional fixed delays}. For suppressed updates, RACS and AS mitigate this attack; the trusted server in RACS will conventionally employ dead reckoning to compensate for any missing packets, so cheaters cannot gain an advantage. A cheater in AS is unable to receive packets from other players if he does not update his decision. In Majority Voting-based schemes, if a cheater intentionally drops the outgoing packets, the honest player may not be able to recover the result if \(k\) is small enough. With respect to fixed delays, the solutions of AS and Majority Voting are the same as the ones to suppressed updates. For RACS, a cheater who attempts a fixed delay triggers the refereeing mechanism, which can detect foul play. With respect to \emph{collusion attacks}, RACS and AS are unable to prevent this attack while Majority Voting mitigates this to an extent. However, Majority Voting can only prevent collusion if the number of cheaters is smaller than \(n-k\), where $n$ is the total number of players and \(k\) is a threshold number of players ($k \leq n$).

\begin{table}[]
\caption{Comparing related work.}
\resizebox{\columnwidth}{!}{%
\begin{tabular}{|rr|c|c|c|c|c|c|c|}
\hline
\multicolumn{2}{|c|}{Cheats}                                                & C/S & Lockstep & AS & PP-CA & RACS & \makecell{Majority\\Vote} & \makecell{\textbf{Our}\\\textbf{Work}} \\ \hline
\multicolumn{1}{|r|}{\multirow{2}{*}{\makecell{Packet\\Forgery}}} & Modification &                 \cmark         &     \cmark &      \cmark                                                                 &        \cmark            & \cmark                &  Weak                                                                    &          \cmark           \\ \cline{2-9} 
\multicolumn{1}{|r|}{}                                & Inconsistency       &       \cmark                      &     \xmark         &          \xmark                                          &       \cmark         & \cmark                                                              &       \cmark                                                                               &       \cmark                                    \\ \hline
\multicolumn{1}{|c|}{\multirow{3}{*}{\makecell{Timing\\Cheats}}}  & \makecell{Suppressed\\Updates}  &        \cmark                     &   \cmark  &          \cmark                                                                &       \cmark &   \cmark                                                                    &  Weak                                                                       &         \cmark                                  \\ \cline{2-9} 
\multicolumn{1}{|r|}{}                                & \makecell{Intentional\\Delay}         &             \cmark                &   \cmark &             \cmark                                                               &           \cmark                                                            & \cmark       &  Weak                                                                       &         \cmark                                  \\ \hline
\multicolumn{2}{|c|}{Collusion}                                             &  \xmark               &   \xmark      &          \xmark                                                                  &       \xmark                                                     &        \xmark       &  \cmark                                                                     & \cmark                          \\ \hline
\end{tabular}%
}

\label{tab:protocol}
\end{table}

\section{Proposed Design}
\label{ch:model}
The previous section examined state-of-the-art methods for tackling cheating in multiplayer online gaming. We note that some protocols may complement others: for instance, RACS (\S\ref{RACS}) is vulnerable to collusion attacks, which can be addressed by Majority Voting (\S\ref{collusion}). The SoI model of AS (\S\ref{AS}) also provides a valuable suggestion for filling the lack of interest model in RACS and CRT. In this section, we propose a framework to address the shortcomings of existing schemes in order to meet the cheats given in Table~\ref{tab:protocol}. 

\subsection{Concepts}

This proposed framework involves three standard entities: a trusted server, $S$; a trusted referee, $R$; and a set of players, $\{P_1, P_2, \dots, P_n\} \in \textbf{P}$, where $n$ is the number of players, with each player having a unique identifier (ID). We note that $S$ and $R$ are controlled by game developers. $S$ has two main components: a back-end database and an authentication facility. The database stores each player's dynamic game states and the long-term static states; it can transmit the data to other entities if necessary. The authentication facility is designed to authenticate players based on their credentials; it will assign an ID to authenticate the player and store functionally necessary information, e.g.\ IP address. $R$ monitors the game state, communicates with players, and enters the verification model if a player is a suspected cheater (e.g.\ does not send responses). The communication and the verification model is described later. Each player, $P_i \in \textbf{P}$, is a human player within the game. A player can receive the game state from $S$ and communicate his gameplay decisions with other players in the game, and $R$. His local client machine is also allowed to compute game events and display the game as it is being played. It is assumed that the number of players is $n \geq 3$, and that there is at least one honest player.

\par  The server, $S$, has independently generated long-term symmetric keys that are shared between each player and $R$. Game time is independent of the player's network conditions. It is broken into rounds of length \(d\), which can be flexibly adjusted in terms of network conditions on \(R\). Game developers should first decide the value of  \(d_{max}\), the maximum amount of time that players can tolerate within one round, and monitor the fluctuation of \(d\) to ensure \(d \leq d_{max}\). A player who fails to update his decision by the end of the round will then be marked as suspicious and trigger the verification model, which will be described later.

\subsection{User Authentication}

\par The first stage is to authenticate a new player and introduce him to the game world. The necessary steps are described as follows:
\begin{enumerate}
\item A new player \(P_A\) sends a login request to the server \(S\) to log in to the game. \(S \) then verifies the identity of \(P_A\) based on the given authentication credentials (e.g. passwords) and records his IP address.
\item If \(P_A\) is successfully authenticated, \(S\) will assess the game state database, and send \(P_A\)'s latest game state, \(X_A\), to \(P_A\)'s host and the referee \(R\), which is encrypted. \(R\) also notifies \(P_A\) of the other online players that are currently involved in the gameplay. Note that \(P_A\) will only be aware of the ID of available players (e.g.\ \(P_B\)) and will not know the game state such as the location of other players in this step.
\item \(R\) broadcasts to other online players that \(P_A\) has joined the server. Similar to above, those players will not learn the game state of \(P_A\). 
\end{enumerate}

\begin{figure}
	\centering
		\includegraphics[scale=.45]{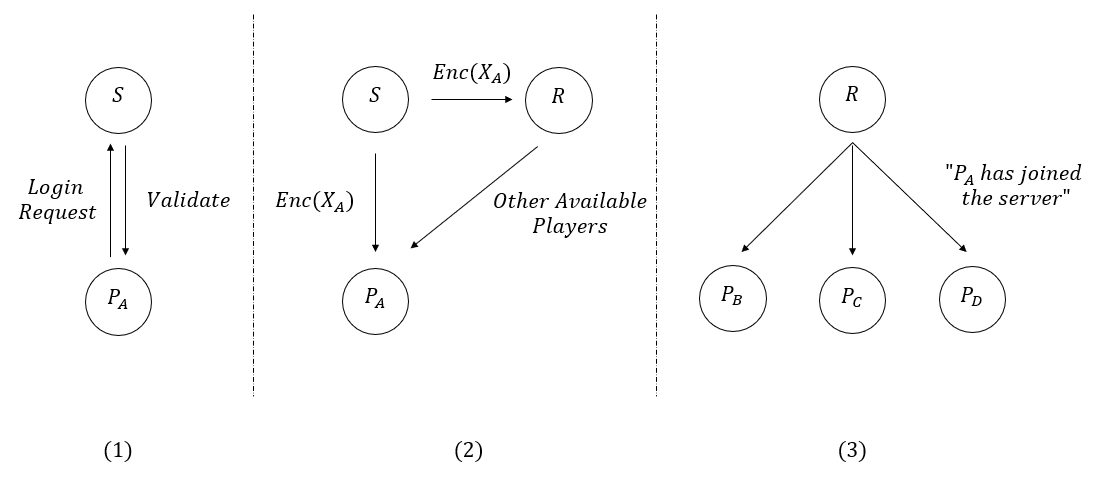}
	\caption{Steps of user authentication.}
	\label{FIG:user_authentication}
\end{figure}

\subsection{Area of Interest}
A player will communicate his game state to other players if and only if they are in his area of interest. The interest model is similar to the SoI model in AS protocol (described in \S\ref{AS}). Throughout the game session, players will continuously update their game states to referee \(R\), who is responsible for managing the interest model. If there is an intersection between players' interest models, \(R\) will inform those players to enter the communication model. The player who is not in any other player's area of interest will continue to send his update to the referee \(R\).
\subsection{Secret Sharing Scheme}
\label{sec:sss}
Our work utilises the Matsumoto-Okabes proposal~\cite{Matsumoto} (described in \S\ref{collusion}), providing additional protection against collusion attacks.  In this model, there is one share dealer, one secret owner, and \(n \in \mathbb{N}\) secret holders. The secret owner first sends his secret \(S\) to the share dealer, who is responsible for allocating a share of the secret to each secret holder. Then, the secret holders perform a calculation based on the received data and send back the results to the secret owner so that he can restore the secret. Considering the possibility of loss, the dealer will decide the threshold value \(k \geq 3\)  so that a secret owner can restore the secret \(S\) given \(k\) shares. 

\par In our proposed framework, the secret \(S\) is the game state. The secret owner \(A\) is a player selected by the referee \(R\) in terms of network conditions, and \(R\) is the share dealer. Note that the secret owner may change throughout the gameplay. If only one player \(B\) is within the SoI of the secret owner, then the secret holders are \(A\), \(B\), and \(R\), and the threshold \(k =3\) (i.e. \(A\) can restore \(S\) if and only if he receives the result from \(A\), \(B\), and \(R\)). Otherwise, the secret holders are \(A\) (the secret owner himself) and the players standing in \(A\)'s SoI. The value of \(k\) is decided by \(R\).

\subsection{Communication}
\par
\begin{figure}[h]
	\centering
		\includegraphics[scale=.75]{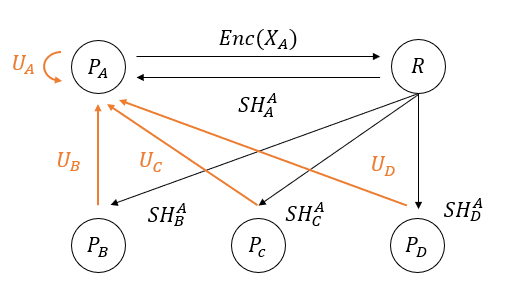}
	\caption{Communication process for $n>2$ players.}
	\label{FIG:comm_four}
\end{figure}

 We first discuss the situation when only two players' SoI models intersect. Since we assume $n \geq 3$ players, this naturally means that there is at least one player who is not involved in this communication. The process is illustrated in Figure \ref{FIG:comm_two}. Without loss of generality, we assume \(P_A\) is the secret owner:
\begin{figure}[h]
	\centering
		\includegraphics[scale=.75]{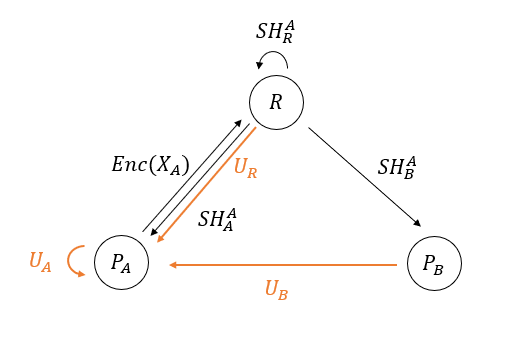}
	\caption{Communication process for $n=2$ players.}
	\label{FIG:comm_two}
\end{figure}
\begin{enumerate}
\item \(P_A\) sends his encrypted game state \(Enc(X_A)\) to \(R\).
\item \(R\) sends a share of the data of \(P_A\) to itself, \(P_A\), and the player in \(P_A\)'s SoI (i.e.\ \(P_B\)). We denote this as \(R\) sending \(SH^A_A, SH^A_B, SH^A_R\) to \(P_A, P_B \), and \(R\) respectively.
\item Based on the share of the data, each share holder (including \(R\)) calculates events and sends back the results \(U_i\) to \(P_A\). Note that Player \(B\) also digitally signs his output and appends the signature to the original message. It is assumed that the verification key of \(P_B\)'s digital signature is publicly available.
\item \(P_A\) restores the results from shares not less than the threshold \(k\); in this scenario, \(k=3\).
\end{enumerate}
The situation when more than two players are communicating is similar (e.g.\ as in Figure~\ref{FIG:comm_four}). While the secret owner and the share dealer remain the same, the secret holders only include players (i.e.\ \(P_A\), \(P_B\), \(P_C\), and \(P_D\)). The value of threshold \(k\) can be equal to either 3 or 4 in this example. Therefore, if one player does not send his packet or sends an incorrect packet to \(P_A\) intentionally or unintentionally, it is still possible for \(P_A\) to restore the results. 

\subsection{Simulation and Verification}
\label{validation}
\begin{figure}
	\centering
		\includegraphics[scale=.75]{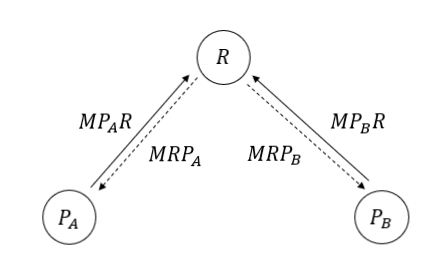}
	\caption{Verification process for $n=2$ players.}
	\label{FIG:comm_two2}
\end{figure}

We assume that a simulation process is executing in parallel to the communication process. At the beginning of a new round, each player is required to send the output he calculated and the data he received in the last round to \(R\) to simulate the gameplay. If \(P_A\) cannot restore the results, he will report that to \(R\) and trigger the verification model. Similar to the communication model, we first assume the scenario where only two players are in the conversation. A diagram of the procedure is given in Figure~\ref{FIG:comm_two2}.

\begin{figure}
	\centering
		\includegraphics[scale=.75]{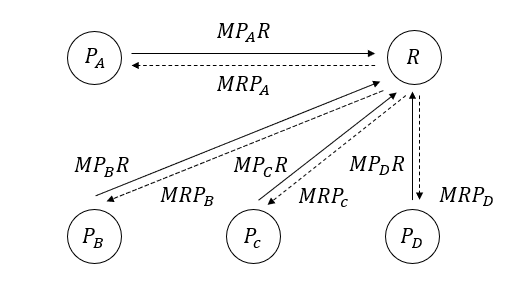}
	\caption{Verification process for $n>2$ players.}
	\label{FIG:comm_42}
\end{figure}

\par After receiving the output from \(P_B\) and \(R\), \(P_A\) will send a message \(MP_AR\) to \(R\). The message comprises \(U_A\), which is the output calculated by \(P_A\); and \(T_A\), which is a summary of the message he received in the last round. The set \(T_A\) includes the ID of sender \(B\) and the digital signature of the message (\(D(U_B)\)). In other words, \(MP_AR = \{U_A, T_A\}\), where \(T_A\) = \(\{B, D(U_B)\}\). Since \(R\) is a trusted entity, we assume its correctness and thus \(U_R\) is not included in the set \(T_A\). The purpose of \(T_A\) is to help \(R\) detect cheaters and inconsistency if any suspicious behavior is reported by \(P_A\), which we will describe in the verification model. With regards to \(P_B\), he only sends his calculated output \(U_B\) to \(R\) since the message he received in the last round  (\(SH_B\)) was from \(R\), which we assume it will be always correct. 
\par If \(P_A\) can restore the results, the game continues and \(R\) remains silent. However, if \(P_A\) is unsuccessful, then \(R\) will enter the verification model and communicate with the player. The solutions to different unsuccessful cases are described as follows:
\begin{enumerate}
\item \(P_A\) reported that he did not receive any message from \(P_B\). He is only able to send \(U_A\) to \(R\). If \(R\) receives the \(U_B\) from \(P_B\), \(R\) will forward the message packet to \(A\) so that the game can continue. If \(R\) does not receive the packet either, he will dead-reckon  \(P_B\)'s avatar and send back the result \(U_B\) to \(A\). He will simultaneously send a message \(MRP_B\) to notify \(P_B\) that none of them have successfully received his message. If \(P_B\) has no response for \(m\) consecutive rounds (the value of \(m\) is defined by \(R\)), \(R\) will consider him as a disconnected player and trigger the leaving model.
\item \(P_A\) reported that he cannot restore the result from the message packets (i.e.\ \(P_B\) sent an incorrect packet). \(R\) will then use \(P_B\)'s public key to verify the digital signature and compare the extracted message with the output sent by \(P_B\). Since \(P_A\) does not know \(P_B\)'s private key, he is not able to forge \(P_B\)'s digital signature. If two outputs are inconsistent, \(R\) will send the correct output \(U_B\) to \(A\) and mark \(B\) as a suspicious player. If two outputs are consistent, it means that \(B\) is wrongly accused. Thus, \(R\) will ignore the report and mark \(A\) as a suspicious player. If the suspicious player is dishonest for \(n\) consecutive rounds (similarly, the value of \(n\) is defined by \(R\)), \(R\) will consider him a cheater and remove him from the game. 
\end{enumerate}

The processes for more than two players, illustrated in Figure~\ref{FIG:comm_42}, are the same as the one described above, i.e.\ the role of \(P_C\) and \(P_D\) is the same as \(P_B\).

\subsection{User Disconnection}
There are two possible leaving models. One such model will be triggered when a player, say \(P_B\), no longer stays in the communication model. The reason can be: 1) he is not in the area of interest of the secret owner; 2) he has been caught cheating; or 3) he has naturally disconnected. \(R\) will send a message \(MRP_B\) to the player to inform him that he has been removed from the conversation. His latest game state will be encrypted and updated to the central server, \(S\).   
\par
\begin{figure}
    \centering
    \includegraphics[scale=.45]{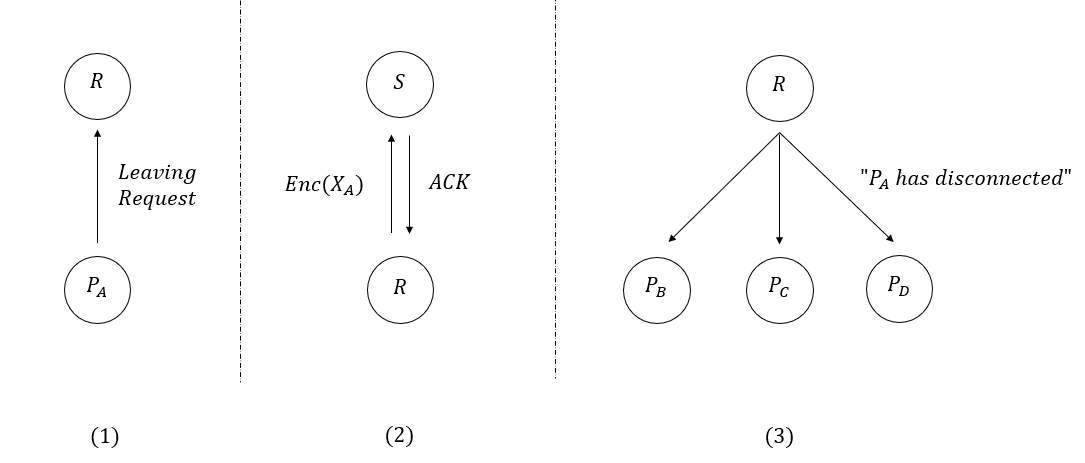}
    \caption{User disconnection process.}
    \label{fig:leaving}
\end{figure}

 Figure \ref{fig:leaving} illustrates the second possible leaving model where a player wants to quit the game. He sends a leaving request to the referee \(R\), who will then update the latest encrypted game state to the server \(S\) and wait for an acknowledgment. Once \(S\) confirms, \(R\) will disconnect the player and broadcast the news to all the other available players. A disconnected player is no longer allowed to communicate with any other entity in this framework.

\section{Evaluation}
\label{ch:evaluation} 
This section evaluates our proposed framework in \S\ref{ch:model}. We employ a two-part evaluation using formal verification using AVISPA and performance simulations.

\subsection{AVISPA Formal Verification}
\label{validationtechnique}
We use AVISPA to verify and evaluate our proposal. The problem of deducing whether a security protocol is susceptible to attack is, in general, undecidable~\cite{OFMC}. Despite this, it is still practical to analyse many real-world security protocols (e.g.\ TLS~\cite{cremers2017comprehensive} and the EMV suite~\cite{basin2021emv}). From \cite{OFMC}: let us assume an initial state, $\mathcal{I}$; a set $\mathcal{R}$ of rules defining the transitions on states; and a set $\mathcal{G}$ of attack rules, or goals, specifying the attack states. Our task is to determine whether an attack state, $g \in \mathcal{G}$, is reachable from $\mathcal{I}$ given in the protocol definition.  

 AVISPA compiles a protocol specification written in the High-Level Protocol Specification Language (HLPSL)~\cite{vigano2006automated} into an  Intermediate Format (IF). The IF is a transition system of infinite-state, which can be analysed by one of four back-ends: \textit{On-the-Fly Model-Checker} (OFMC) \cite{OFMC}, \textit{CL-based Attack Searcher} (CL-AtSe) \cite{CL-ATSE}, \textit{SAT-based Model Checker} (SATMC) \cite{satmc} and \textit{Tree Automata-based Protocol Analyzer} (TA4SP) \cite{boichut2009tree} for identifying valid attack states from a protocol's definition. Upon termination, the back-end generates a result of analysis stating whether an attack was found against the input protocol.
 
 \begin{figure}
    \centering
    \includegraphics[width=0.9\linewidth]{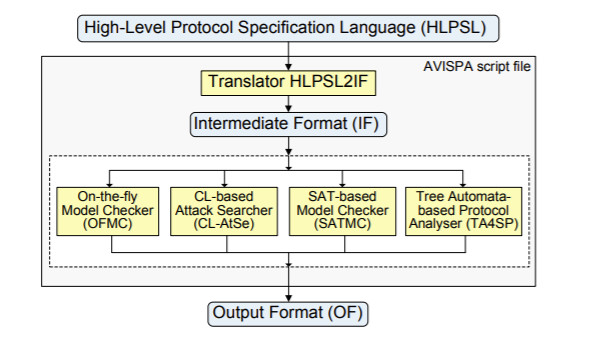}
    \caption{High-level workflow of AVISPA~\cite{vigano2006automated}.} 
    \label{fig:AVISPA_architecture}
\end{figure}

The following security goals can be assessed. \emph{Message confidentiality}: written using the {\tt secrecy\_of <identifier>} in the HLPSL. Note that {\tt
<identifier>} is the message in question. To verify this security goal, an intruder that is out of the agent set will attempt to learn the information. If successful, then the security goal is failed.

\emph{Strong authentication}: of an agent \(X\) by another agent \(Y\). Two predicates {\tt witness} and {\tt request} are used. The {\tt witness} predicate is written in the role of the authenticated agent \(X\). It declares that \(X\) is a witness of the information, where the {\tt request} predicate is written in the role of agent \(Y\) who does the authentication. It declares that \(Y\) requires a check of the information. The {\tt witness} predicate must be emitted before the  {\tt request} predicate.  This security goal is violated if an intruder other than the agent \(Y\) emits a {\tt request} event, or no valid {\tt witness} event is found before the {\tt request} event. 

\emph{Weak authentication}: of an agent \(X\) by another agent \(Y\). This goal is similar to strong authentication, except that it does not check for replay attacks.

\subsection{Protocol Scope}
\label{sec:protocol-scope}
The proposed framework consists of seven models: \ballnumber{1} the joining and \ballnumber{2} and leaving models, which are parts of a traditional authentication process; \ballnumber{3} the Area of Interest model and \ballnumber{4} secret sharing process are adapted previous research, i.e.~\cite{Matsumoto}; \ballnumber{5} the simulation and \ballnumber{6} verification models involve communication between the player and a trusted entity. It is \ballnumber{7} the communication model that represents our principal contribution. This model assumes two players (i.e.\ \(A\) and \(B\)) and a trusted referee \(R\). 
\(R\) has generated a shared long-term symmetric key set \(K_{AR}\) (\(K_{BR}\) ) with Player \(A\) (\(B\)). The symmetric keys are safely stored in a separate location where the attacker cannot overwrite them. \(A\) and \(B\) do not share any secrets or any symmetric keys before the protocol starts. Due to network conditions, \(A\) is selected as the supernode and thus the calculation is based on his game state. 
\par The output \(U_B\) is assumed to be signed by player \(B\), and \(A\) and \(R\) have learned the verification key in advance. Thus, if \(A\) claims that he cannot recover the secret from the events he received, he can then send the digital signature to \(R\) and trigger the verification model for future investigation. As \(R\) is a trusted entity, it is assumed that he will not modify his output \(U_R\) and send an incorrect result to \(A\).

\subsection{Proposed Communication Protocol}
\label{proposedcommunicationprotocol}
The exchanges within our proposed communication protocol is shown in Figure~\ref{fig:AVISPA_msg}, while Table \ref{table:AVISPA_notation} presents the notation in use. This model can be considered as a sequence of a key agreement (steps 1--4) and a secret sharing (steps 5--9) protocol. These steps are described as follows:

\begin{table}
\caption{Protocol notation.}
\begin{tabularx}{0.4\textwidth}{@{}llll@{}}
\toprule
 \textbf{Notations }&\textbf{Meaning} \\
 \hline
  \(A\),\(B\),\(R\) & Communicating participants\\
  
   \(N_A\) & Nonce generated by \(A\) \\

   \(X_A\) & The game state of \(A\) \\

   \(R_A\) & Request sent by \(A\) \\
   \(SH_A\) &The share of the secret for \(A\) \\

 \(U_B\) & The calculated output of \(B\) \\
  
    \(H(K_{AR},...)\)   &  An HMAC function under key \(K_{AR}\)  \\
     
  \(K_{AR}\)& Symmetry key shared between \(A\) and \(R\).   \\

   \(
\{K_{AB}\}K_{AR} \) & Message \(K_{AB}\) is encrypted under key \(K_{AR}\)\\

\bottomrule
\end{tabularx}
\label{table:AVISPA_notation}
\end{table}

\begin{enumerate}
\item Before the protocol starts, \(A\) and \(B\) are informed by \(R\) that their areas of interest have intersected. \(A\) then sends a communication request to \(B\), comprising a fresh nonce \(N_A\) and his ID, \(A\). 
\item \(B\) sends a key establishment request to the trusted referee \(R\) since \(A\), \(B\) do not share any symmetric key.  The request involves the received nonce  \(N_A\), a freshly generated nonce \(N_B\), and his ID \(B\), alongside a MAC of the set of messages\(\{N_A,N_B,A,B\}\) using \(K_{BR}\).
\item[3--4.] \(R\) generates a session key, \(K_{AB}\), and sends it to \(A\) and \(B\). \(R\) encrypts the session key  \(K_{AB}\) with the symmetric key \(K_{AR}\) (\(K_{BR}\)), the ID of \(B\) (\(A\) ), and a MAC over \(\{N_A,B,\{K_{AB}\}K_{AR}\}\) ( \(\{N_B,A,\{K_{AB}\}K_{BR}\}\) ) using the long-term key \(K_{AR}\) (\(K_{BR}\)). 
\item[5.] After completing the key agreement protocol, $A$ sends a request \(R_A\) to \(R\) for processing his game state.  A concatenation of request \(R_A\) and game state \(X_A\), encrypted with the key \(K_{AR}\)  is sent alongside the request. Since the game state must not be accessed by entities other than \(A\) and \(R\), this communication channel is read and modification protected. This is achieved by a dynamic one-time symmetric key, selecting from a symmetric key set owned by \(A\) and \(R\).
\item[6.] Similar to step 5, \(B\) also sends a request \(R_B\) to \(R\), alongside his ID, \(B\).
\item[7.] From \(X_A\), \(R\) prepares 3 shares of secret \(SH_A,SH_B,\) and \(SH_R\) for \(A,B\), and \(R\) respectively. \(R\) calculates the output \(U_R\) using his share of secret \(SH_R\), returning the response to \(A\). The reply includes the share of the secret for \(A\), the output \(U_R\) from \(R\), and a corresponding MAC function. Note that the request \(R_A\) is also included in the MAC function for message authentication.
\item[8.] Concurrently, \(R\) sends the secret \(SH_B\) to \(B\), alongside the ID \(A\) to inform \(B\) to whom the secret belongs. A MAC of the message is also attached and encrypted with \(K_{BR} \). 
\item[9.] \(B\) calculates the event and sends it back to \(A\). The event \(U_B\) is encrypted with the fresh session key \(K_{AB}\), and the MAC function includes the nonce \(N_A\), which is sent from \(A\) in step 1.
\end{enumerate}

\begin{figure*}%
\centering
\captionsetup{justification=centering}
\scalebox{0.9}{
\fbox{
\procedure{Proposed protocol (2-player setting)}{
 \textbf{Player \(A\)} \< \<\textbf{Player \(B\)}  \< \<\textbf{Referee \(R\)}  \\
 \<\sendmessageright{length=5
 cm,top=\text{\(N_A\), \(A\)}} \\
 \<\<\<\sendmessageright{length=5cm,top=\text{\(N_A\), \(N_B\), \(B\), \(H(K_{BR},N_A,N_B,A,B)\)}} \\
  \<\sendmessageleftx[14cm]{12}{\text{\(\{K_{AB}\}K_{AR}\), \(B\),\(H(K_{AR},N_A,B,\{K_{AB}\}K_{AR})\)}}\<  \\
  \<\<\<\sendmessageleft{length=5cm,top=\text{\(\{K_{AB}\}K_{BR}\), \(A\),\(H(K_{BR},N_B,A,\{K_{AB}\}K_{BR})\)}} \\
   \<\sendmessagerightx[9cm]{12}{\text{\(R_A\),\(\{R_A,X_A\}K_{AR}\)}}\< 
   \\
    \<\<\<\sendmessageright{length=5cm,top=\text{\(R_B\), \(B\)}} \\
      \<\sendmessageleftx[9cm]{12}{\text{\(\{SH_A\}K_{AR}\), \(\{U_R\}\}K_{AR}\),\(H(K_{AR},R_A,B,\{SH_A\}K_{AR},\{U_R\}K_{AR})\)}}\<  \\
          \<\<\<\sendmessageleft{length=5cm,top=\text{\(\{SH_B\}K_{BR}\),A,\(H(K_{BR},R_B,A,\{SH_B\}K_{BR},\)}} \\
           \<\sendmessageleft{length=5cm,top=\text{\(\{U_B\}K_{AB}\), \(H(K_{AB},N_A,\{U_B\}K_{AB})\)}} \\
}}}
\caption{Proposed protocol.}   \label{fig:AVISPA_msg}
\end{figure*}

\subsection{Model Evaluation}
\label{protocolgoal}
We test our protocol against the three security properties available through AVISPA: secrecy, authentication, and weak authentication. Specifically, we verify the following security properties of the following assets, which serve as valuable targets during online gaming scenarios:

\begin{enumerate}
\item \textbf{\(X_A\) secrecy}. The protocol should guarantee that game state \(X_A\) can only be accessed by the owner \(A\) and the referee \(R\). If another entity is able to compromise the confidentiality and integrity of \(X_A\), he can gain additional information about \(A\). Moreover, he may be able to modify \(X_A\) so that \(R\) will falsely accuse \(A\) of cheating and remove \(A\) from the game. This property is verified using the statement {\fontfamily{qcr}\selectfont
secrecy\_of xa}.
\item \textbf{\(U_R\) secrecy}. The protocol should ensure that only \(A\) and \(R\) are able to learn the calculated output \(U_R\). \(A\) will always expect the correctness of \(U_R\) since it is from the trusted referee \(R\). If an attacker tampers \(U_R\), \(A\) can never restore the secret. Note that this security goal is omitted if more than two players are involved in the communication model. This property is verified using the statement {\fontfamily{qcr}\selectfont
secrecy\_of ur}.
\item \textbf{\(U_B\) secrecy}. The protocol should guarantee that event \(U_B\) can only be accessed by the owner \(B\), the recipient \(A\), and the referee \(R\). If \(U_B\) is tampered with, the gameplay will be disrupted and \(B\) will be mistakenly accused of cheating. To verify this property, the statement  {\fontfamily{qcr}\selectfont
secrecy\_of ub} is introduced in the protocol specification.
\item \textbf{\(K_{AB}\) secrecy}. \(K_{AB}\) is generated for \(A\) and \(B\) by \(R\). Therefore, only these three entities are allowed to learn the value of \(K_AB\). This goal is achieved by the statement  {\fontfamily{qcr}\selectfont
secrecy\_of kab}.
\item The \textbf{strong authentication} of \(R\) by \(B\) through the nonce \(N_B\) to be secure against the man-in-the-middle (MITM) attack. To verify this security goal, {\fontfamily{qcr}\selectfont
witness}and {\fontfamily{qcr}\selectfont
request} statements are also included  in the role specifications of \(B\) and \(R\) respectively.

\item The \textbf{strong authentication} of \(B\) by \(A\) through the nonce \(N_B\) to be secure against a MITM attack. To verify this security goal, {\fontfamily{qcr}\selectfont
witness }statement is added in the role specification of \(A\).
 \end{enumerate}


AVISPA uses a Dolev-Yao adversarial model \cite{DY}, denoted by {\fontfamily{qcr}\selectfont
channel(dy)} in the roles as well as an intruder model. An active DY intruder \(i\) is able to access and analyse the message in sending channel {\fontfamily{qcr}\selectfont
  SND(dy)} and modify the incoming message in receiving channel  {\fontfamily{qcr}\selectfont RCV(dy)}. In this simulation, it is assumed that an intruder could compromise either \(A\) and \(B\) and learn the long-term symmetric key shared with \(R\). Since \(R\) is a trusted entity, the intruder is not able to play the role of \(R\). Therefore, three different sessions are instantiated in the HLPSL specification: (1) a session among \(A\),\(B\) and \(R\), (2) a session among \(i\), \(B\) and \(R\) and (3) a session among \(A\), \(i\), and \(R\). The Security Protocol Animator (SPAN) is used to simulate the HLPSL specification file. It can interactively build a Message Sequence Chart (MSC) of the simulated protocol and automatically generate attacks MSC using the AVISPA verification tools.  We can report that AVISPA found no attacks against our proposed protocol using all four backends.


\setlength{\emergencystretch}{3em}


\subsection{Informal Analysis}

\label{evaluation2}

Let us consider the proposed protocol against online game-specific security issues. Throughout this discussion, we assume that \(P_A\) is an honest player while \(P_B\) is a cheater.  We note that, to prevent packet modification attacks, if the cheater modifies his network packets, the honest player cannot restore the secret from his output \(U_C\). Thus, his tampered packets will be withdrawn. Based on the result from the honest player, referee \(R\) will simulate the gameplay for the cheater, making packet forgery impossible to achieve. We posit that inconsistency cheats are inherently mitigated as players will not send the same packet to different players concurrently.
For timing cheats, both suppressed updates and fixed delays are addressed with the help of the verification model. If the cheater intentionally drops or delays his outgoing packets, then the protocol will enter the case that \(P_A\) reported that he did not receive any message from \(P_B\)', which was described in \S\ref{validation}.
For collusion detection, similar to Majority Voting protocols, this protocol has a strong performance against this attack if the number of cheaters does not exceed \(n-k\).

\section{Simulation}
\label{ch:simulation}

We now describe results of the proposed protocol in Python. In this simulation, we implemented all 7 models that comprise a realistic gaming communication scenario given in \S\ref{sec:protocol-scope}. We implement each model in Python, including our proposed protocol, using simulations based on 10--10K runs of each model over a local area network (LAN). Experimental hardware included a consumer PC with 32GB RAM, Intel Core i5-4690K CPU (3.9GHz), running Ubuntu 24.10 and Python 3.12.3 using the \texttt{cryptography} and \texttt{shamirs} packages.

\subsection{User Connection}

To enter the game world, a player \(P_A\) is required to enter his login credentials. Once the login credentials are verified, he will subsequently receive his game state. In the simulation, the game state is represented by a randomly large (128 bits sized) integer.  Table \ref{tab:user-authen-server}  and \ref{tab:user-authen-client1} shows the average execution time and the physical memory usage for the server and player \(P_A\) respectively. For both server and player \(P_A\), as the number of runs increases, the average execution time decreases and tends to be around 0.003 seconds. The physical memory usage remains around 19 to 20 Megabytes regardless of the number of runs.

\begin{table}[]
\resizebox{\columnwidth}{!}{%
\begin{tabular}{|l|l|l|l|l|}
\hline
\textbf{Number of runs}                               & 10            & 100           & 1,000         & 10,000        \\ \hline
\textbf{\begin{tabular}[c]{@{}l@{}}Average execution time (ms)\end{tabular}}         & 8.1        & 4.2        & 3.1        & 3.3        \\ \hline
\textbf{\begin{tabular}[c]{@{}l@{}}Average memory usage (MB) \end{tabular}} & 19.34 & 19.33 & 19.08 & 19.41 \\ \hline
\end{tabular}%
}
\caption{User authentication process -- result for the server.}
\label{tab:user-authen-server}
\end{table}
\hfill

\begin{table}[]
\resizebox{\columnwidth}{!}{%
\begin{tabular}{|l|l|l|l|l|}
\hline
\textbf{Number of runs}                               & 10            & 100           & 1,000         & 10,000        \\ \hline
\textbf{\begin{tabular}[c]{@{}l@{}}Average execution time (ms)\end{tabular}}          & 8.8        & 4.2        & 3.5        & 3.4        \\ \hline
\textbf{\begin{tabular}[c]{@{}l@{}}Average memory usage (MB)\end{tabular}} & 19.67 & 19.92 & 20.20 & 20.62 \\ \hline
\end{tabular}%
}
\caption{User authentication process -- result for the player \(P_A\).}
\label{tab:user-authen-client1}
\end{table}
\subsection{Area of Interest}

Once \(P_A\) is in the game world, he will constantly send out his location message to the server. The server will confirm the connection by sending out the message \lq Connection established. Searching for nearby players...\rq  and notify \(P_A\) that there is another player \(P_B\)  in his area of interest. In the simulation, \(P_A\) and \(P_B\) will send a random integer to the referee \(R\) respectively. \(R\) will subsequently calculate the difference between two integers to decide if two players are nearby. Once the difference is below a given threshold, \(R\) will send out a message \lq Nearby Player identified.\rq  to inform both players.

\par  Table \ref{tab:AoI-server}, \ref{tab:AoI-client1} and \ref{tab:AoI-client2} show the average execution time and the physical memory usage for \(R\), \(P_A\), and \(P_B\)  respectively. It can be seen that the average execution time remained around 1 second regardless of the number of runs for each node in the model. The average physical memory usage fluctuated, but the spike was eliminated after running 10,000 times.

\begin{table}[]
\resizebox{\columnwidth}{!}{%
\begin{tabular}{|l|l|l|l|l|}
\hline
\textbf{Number of runs}                               & 10            & 100           & 1,000         & 10,000        \\ \hline
\textbf{\begin{tabular}[c]{@{}l@{}}Average execution time (ms)\end{tabular}}         & 1001        & 1001        & 1002        & 1002       \\ \hline
\textbf{\begin{tabular}[c]{@{}l@{}}Average memory usage (MB)\end{tabular}} & 19.36 & 39.01 & 19.71 & 17.94 \\ \hline
\end{tabular}%
}
\caption{Area of Interest process -- result for the referee \(R\).}
\label{tab:AoI-server}
\end{table}
\hfill

\begin{table}[]
\resizebox{\columnwidth}{!}{%
\begin{tabular}{|l|l|l|l|l|}
\hline
\textbf{Number of runs}                               & 10            & 100           & 1,000         & 10,000        \\ \hline
\textbf{\begin{tabular}[c]{@{}l@{}}Average execution time (ms)\end{tabular}}          & 1002        & 1002         & 1003        & 1003        \\ \hline
\textbf{\begin{tabular}[c]{@{}l@{}}Average memory usage (MB)\end{tabular}} & 20.22 & 19.92 & 41.50 & 19.04 \\ \hline
\end{tabular}%
}
\caption{Area of Interest process -- result for the player \(P_A\).}
\label{tab:AoI-client1}
\end{table}

\begin{table}[]
\resizebox{\columnwidth}{!}{%
\begin{tabular}{|l|l|l|l|l|}
\hline
\textbf{Number of runs}                               & 10            & 100           & 1,000         & 10,000        \\ \hline
\textbf{\begin{tabular}[c]{@{}l@{}}Average execution time (ms)\end{tabular}}          & 1001        & 1001        & 1001        & 1001        \\ \hline
\textbf{\begin{tabular}[c]{@{}l@{}}Average memory usage (MB)\end{tabular}} & 28.67 & 28.63 & 22.78 & 28.92 \\ \hline
\end{tabular}%
}
\caption{Area of Interest process -- result for the player \(P_B\).}
\label{tab:AoI-client2}
\end{table}

 \subsection{Communication}
\begin{figure}[]

     \pgfplotstableread{ 
Label     AverageLength
M9   144
M8   143.342  
M7    222.69 
M6     21 
M5 90 
M4   141 
M3    141 
M2     142 
M1 36
    }\testdata

    \begin{tikzpicture}

    \begin{axis}[
    title={Average Length of Message},
    xbar stacked,   
    xmin=0,         
    ytick=data,     
    legend style={at={(axis cs:65,0.2)},anchor=south west},
    yticklabels from table={\testdata}{Label}  
    ]

    \addplot [fill=gray!60,    point meta=x,
    nodes near coords,
    nodes near coords align={anchor=east},
    every node near coord/.append style={
        black,
        fill=gray,
        fill opacity=0,
        text opacity=1,
        outer sep=\pgflinewidth 
    }
    ] table [x=AverageLength, meta=Label,y expr=\coordindex] {\testdata};   

    \end{axis}
    \end{tikzpicture}
    \caption{Average message length (bytes). }
    \label{tab:communication}
\end{figure}
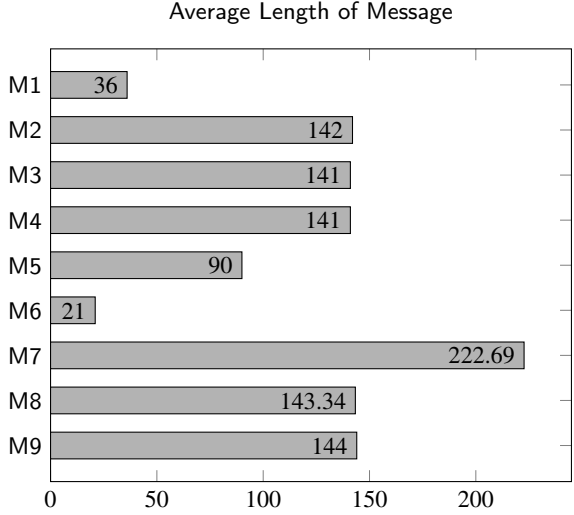

\begin{figure}[]

     \pgfplotstableread{ 
Label     AverageLength
M9   53.21
M8    63.08 
M7    65.77 
M6    60.90 
M5    62.13 
M4    52.65 
M3    62.38 
M2    61.00
M1    50.41
    }\testdata

    \begin{tikzpicture}

    \begin{axis}[
    title={Average Message Transmission Time},
    xbar stacked,   
    xmin=0,         
    ytick=data,     
    legend style={at={(axis cs:65,0.2)},anchor=south west},
    yticklabels from table={\testdata}{Label}  
    ]

    \addplot [fill=gray!60,    point meta=x,
    nodes near coords,
    nodes near coords align={anchor=east},
    every node near coord/.append style={
        black,
        fill=gray,
        fill opacity=0,
        text opacity=1,
        outer sep=\pgflinewidth 
    }
    ] table [x=AverageLength, meta=Label,y expr=\coordindex] {\testdata};   

    \end{axis}
    \end{tikzpicture}
    \caption{Average message transmission times (ms).}
    \label{tab:communication-times}
\end{figure}
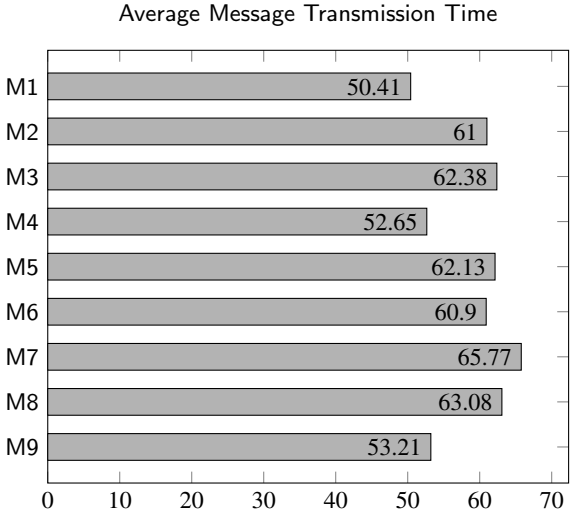

 \par We implemented the communication model described in \S\ref{proposedcommunicationprotocol}. Table \ref{tab:communication} shows the average length of each message after re-running the communication model 10,000 times. It is assumed that each nonce number and key is 32 bytes. It can be seen that message 7 has the largest packet size (222.69 bytes). This is because a share of secret \(SH_A\) and the output \(U_R\) from \(R\) are included in the message. Messages 2, 3, 4, 8, and 9 have similar packet sizes (around 142 bytes) as the structure of those messages is highly similar. The length of message 5 is 90 bytes as MAC is not included in the message. Lastly, Message 1 and Message 6 follow a similar structure, but Message 1 is longer as it includes a nonce number instead of a request \(R_B\) in Message 6. We also give average transmission times in Figure~\ref{tab:communication-times} averaged over 10,000 protocol instances. The total message exchange time took $\sim$500 milliseconds ($\mu=541.53ms$, $\sigma=16.88ms$).

\subsection{User Disconnection}
 
If a player wants to disconnect the game world, he sends a disconnection request to the server. His game state, represented by a randomly large (128 bits sized) integer, is also sent across to the server. He will lose his access to the game world unless he goes through the user authentication process again. Table \ref{tab:user-dc-server}  and \ref{tab:user-dc-client1} show the average execution time and the physical memory usage for the server and player \(P_A\) respectively. For both server and player \(P_A\), as the number of runs increases, the average execution time decreases and tends to be below 0.001 seconds. The memory usage remains around 19--20MB regardless of the number of runs.

\begin{table}[]
\resizebox{\columnwidth}{!}{%
\begin{tabular}{|l|l|l|l|l|}
\hline
\textbf{Number of runs}                               & 10            & 100           & 1,000         & 10,000        \\ \hline
\textbf{\begin{tabular}[c]{@{}l@{}}Average execution time (ms)\end{tabular}}          & 53.3        & 51.1        & 0.2        & 0.1        \\ \hline
\textbf{\begin{tabular}[c]{@{}l@{}}Average memory usage (MB)\end{tabular}} & 19.32 & 19.37 & 19.42 & 19.83 \\ \hline
\end{tabular}%
}
\caption{User disconnection process -- result for the server.}
\label{tab:user-dc-server}
\end{table}

\begin{table}[]
\resizebox{\columnwidth}{!}{%
\begin{tabular}{|l|l|l|l|l|}
\hline
\textbf{Number of runs}                               & 10            & 100           & 1,000         & 10,000        \\ \hline
\textbf{\begin{tabular}[c]{@{}l@{}}Average execution time (ms)\end{tabular}}          & 52.0        & 51.7        & 0.3        & 0.2        \\ \hline
\textbf{\begin{tabular}[c]{@{}l@{}}Average memory usage (MB)\end{tabular}} & 20.05 & 20.34 & 19.78 & 20.31 \\ \hline
\end{tabular}%
}
\caption{User disconnection process -- result for the player \(P_A\).}
\label{tab:user-dc-client1}
\end{table}

\section{Conclusion}

In this work, we examined the state of the art in existing cheat protection proposals against network packet-level attacks.  To bridge the limitations of existing protocols, which tend to be fragmented in terms of design goals, we designed, implemented and evaluated a new protocol for detecting packet cheats, timing cheats, and in-game collusion in multiplayer games.
 Through conducting this work, some limitations and potential areas of research were identified. Most notably, our work relies on a single trusted referee, which can represent a single point of failure. If the referee is unable to function, it will critically impact the gameplay from working as intended. To improve our protocol's resilience and scalability, it may therefore be desirable to distribute the referee's functions among several referees, who coordinate their decisions. Moreover, it would be useful to implement and evaluate the proposed protocol in a real-world game with human players in order to assess its effectiveness, and to explore the application of post-quantum cryptography as a future extension. In conclusion, it is hoped that the main findings of this paper can enhance the understanding of online game cheating and inspire future research in this area. 
\Urlmuskip=0mu plus 1mu\relax
\footnotesize
\bibliographystyle{ieeetr}
\bibliography{cas-refs}

\begin{thebibliography}{10}

\bibitem{Statista_2018}
Statista, ``Digital video game market revenue worldwide from 2017 to 2027.'' \url{https://www.statista.com/statistics/1344686/global-digital-gaming-revenue/}, 2023.

\bibitem{granados_2018}
N.~Granados, ``Report: Cheating is becoming a big problem in online gaming.'' \url{https://www.forbes.com/sites/nelsongranados/2018/04/30/report-cheating-is-becoming-a-big-problem-in-online-gaming/\#5f330df77663}, 2018.

\bibitem{racs}
S.~Webb, S.~Soh, and W.~Lau, ``{RACS}: A referee anti-cheat scheme for {P2P} gaming,'' in {\em 17th Int'l Workshop on Network and Operating Systems Support for Digital Audio and Video}, pp.~34--42, ACM, 2007.

\bibitem{pp-ca}
J.~D. Pellegrino and C.~Dovrolis, ``Bandwidth requirement and state consistency in three multiplayer game architectures,'' in {\em 2nd Workshop on Network and System Support for Games}, pp.~52--59, 2003.

\bibitem{lockstep}
N.~E. {Baughman}, M.~{Liberatore}, and B.~N. {Levine}, ``Cheat-proof playout for centralized and peer-to-peer gaming,'' {\em IEEE/ACM Transactions on Networking}, Feb 2007.

\bibitem{eac}
{Epic Games}, ``Easy anti-cheat.'' \url{https://www.easy.ac/en-us/}, 2023.

\bibitem{battleye}
{BattlEye Innovations}, ``Battleye.'' \url{https://www.battleye.com/}, 2023.

\bibitem{guigo2014next}
N.~Guigo and J.~S. John, ``Next level cheating and leveling up mitigations,'' {\em Black Hat Europe}, vol.~2014, 2014.

\bibitem{Matsumoto}
K.~Matsumoto and Y.~Okabe, ``A collusion-resilient hybrid {P2P} framework for massively multiplayer online games,'' in {\em IEEE 41st Annual Computer Software and Applications Conference}, IEEE, 2017.

\bibitem{armando2005avispa}
A.~Armando, D.~Basin, Y.~Boichut, Y.~Chevalier, L.~Compagna, J.~Cu{\'e}llar, {\em et~al.}, ``The {AVISPA} tool for the automated validation of internet security protocols and applications,'' in {\em Int'l Conference on Computer Aided Verification}, pp.~281--285, Springer, 2005.

\bibitem{YanandRandell}
J.~Yan and B.~Randell, ``A systematic classification of cheating in online games,'' in {\em 4th ACM SIGCOMM Workshop on Network and System Support for Games}, ACM, 2005.

\bibitem{pritchard_2000}
M.~Pritchard, ``How to hurt the hackers: The scoop on internet cheating and how you can combat it.'' \url{https://www.gamasutra.com/view/feature/131557/how_to_hurt_the_hackers_the_scoop_.php/}, 2000.

\bibitem{Yan}
J.~{Yan}, ``Security design in online games,'' in {\em 19th Annual Computer Security Applications Conference}, pp.~286--295, 2003.

\bibitem{feijoo2012mobile}
C.~Feijoo, J.-L. G{\'o}mez-Barroso, J.-M. Aguado, and S.~Ramos, ``Mobile gaming: Industry challenges and policy implications,'' {\em Telecommunications Policy}, vol.~36, no.~3, pp.~212--221, 2012.

\bibitem{Myeongjae_2019}
L.~Myeongjae, ``Monitoring to prevent game cheating.'' \url{https://engineering.linecorp.com/en/blog/monitoring-to-prevent-game-cheating/}, 2019.

\bibitem{NEO}
C.~GauthierDickey, D.~Zappala, V.~Lo, and J.~Marr, ``Low latency and cheat-proof event ordering for peer-to-peer games,'' in {\em 14th Int'l Workshop on Network and Operating Systems Support for Digital Audio and Video}, ACM, 2004.

\bibitem{Corman}
A.~Corman, S.~Douglas, P.~Schachte, and V.~Teague, ``A secure event agreement {(SEA)} protocol for peer-to-peer games,'' in {\em 1st Int'l Conf. on Availability, Reliability and Security}, IEEE, 2006.

\bibitem{endo2006cheat}
K.~Endo, M.~Kawahara, and Y.~Takahashi, ``Cheat prevention for massively multiplayer online distributed services,'' {\em IPSJ Journal}, vol.~47, no.~4, pp.~1087--1098, 2006.

\bibitem{Aronson_1997}
J.~Aronson, ``Dead reckoning: Latency hiding for networked games.'' \url{https://www.gamasutra.com/view/feature/131638/dead_reckoning_latency_hiding_for_.php}, 1997.

\bibitem{Nichols}
J.~Nichols and M.~Claypool, ``The effects of latency on online madden nfl football,'' pp.~146--151, 01 2004.

\bibitem{article}
A.~Yahyavi and B.~Kemme, ``Peer-to-peer architectures for massively multiplayer online games: A survey,'' {\em ACM Computing Surveys}, vol.~46, 2013.

\bibitem{AS}
N.~E. {Baughman} and B.~N. {Levine}, ``Cheat-proof playout for centralized and distributed online games,'' in {\em 20th IEEE Int'l Conference on Computer Communications}, 2001.

\bibitem{shamir1979share}
A.~Shamir, ``How to share a secret,'' {\em Communications of the ACM}, vol.~22, no.~11, pp.~612--613, 1979.

\bibitem{OFMC}
S.~M{\"o}dersheim and L.~Vigan{\`o}, ``The open-source fixed-point model checker for symbolic analysis of security protocols,'' in {\em Foundations of Security Analysis and Design V}, pp.~166--194, Springer, 2009.

\bibitem{cremers2017comprehensive}
C.~Cremers, M.~Horvat, J.~Hoyland, S.~Scott, and T.~van~der Merwe, ``A comprehensive symbolic analysis of {TLS} 1.3,'' in {\em ACM Conference on Computer and Communications Security}, 2017.

\bibitem{basin2021emv}
D.~Basin, R.~Sasse, and J.~Toro-Pozo, ``The {EMV} standard: Break, fix, verify,'' in {\em IEEE Symposium on Security and Privacy}, IEEE, 2021.

\bibitem{vigano2006automated}
L.~Vigano, ``Automated security protocol analysis with the avispa tool,'' {\em Electronic Notes in Theoretical Computer Science}, vol.~155, pp.~61--86, 2006.

\bibitem{CL-ATSE}
M.~Turuani, ``The {CL-Atse} protocol analyser,'' in {\em Int'l Conference on Rewriting Techniques and Applications}, pp.~277--286, Springer, 2006.

\bibitem{satmc}
A.~Armando, R.~Carbone, and L.~Compagna, ``Satmc: a sat-based model checker for security protocols, business processes, and security apis,'' {\em International Journal on Software Tools for Technology Transfer}, vol.~18, no.~2, pp.~187--204, 2016.

\bibitem{boichut2009tree}
Y.~Boichut, P.-C. H{\'e}am, and O.~Kouchnarenko, ``Tree automata for detecting attacks on protocols with algebraic cryptographic primitives,'' {\em Electronic Notes in Theoretical Computer Science}, 2009.

\bibitem{DY}
D.~Dolev and A.~Yao, ``On the security of public key protocols,'' {\em IEEE Trans. Inf. Theor.}, vol.~29, p.~198–208, Sept. 2006.

\end{thebibliography}


\bio{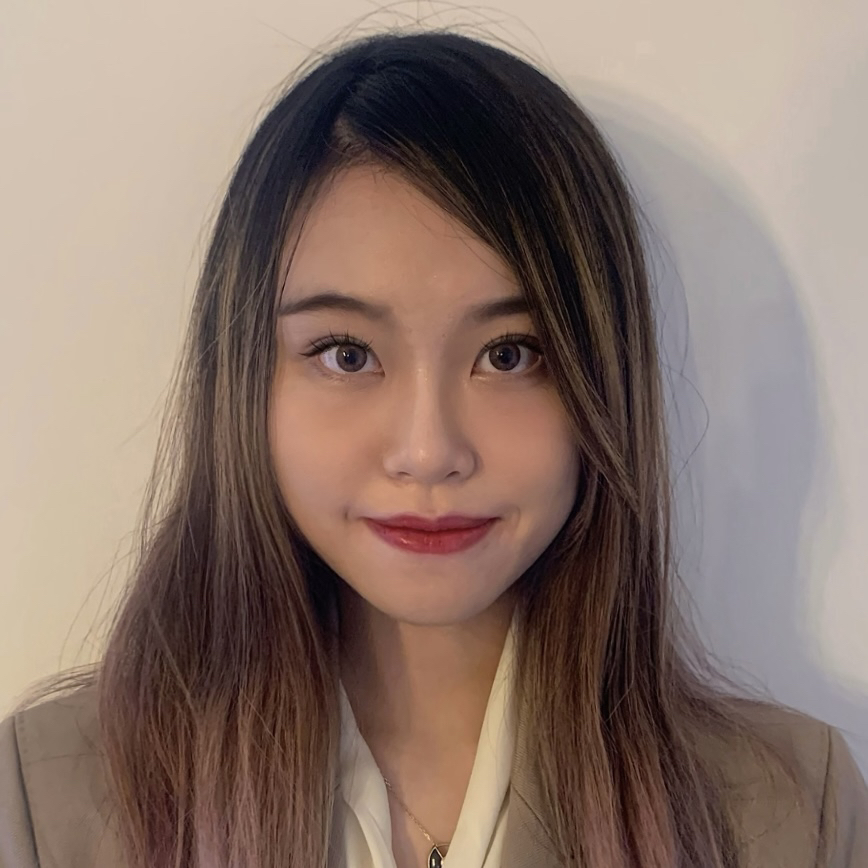}
\textbf{Yaqi Cai} obtained her M.Sc.\ in Information Security in Royal Holloway University of London and her B.Sc.\ in Mathematics with Economics from University College London. Her research interests include game security.
\endbio

\vspace{28pt}
\bio{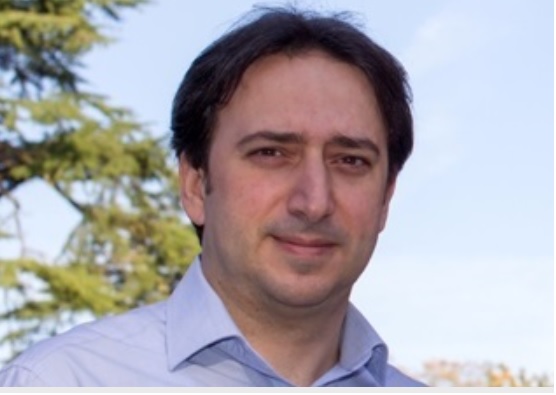}
\textbf{Konstantinos Markantonakis} (B.Sc., M.Sc., M.B.A, Ph.D.) is a Professor of Information Security in Royal Holloway University of London, and the Director of the Smart Card and IoT Security Centre (SCC). He obtained his B.Sc.\ from Lancaster University, and M.Sc., Ph.D.\ and M.B.A.\ from Royal Holloway, University of London. His research interests include smart card security and applications, secure cryptographic protocol design, and embedded systems security.
\endbio
\vspace{-1pt}
\bio{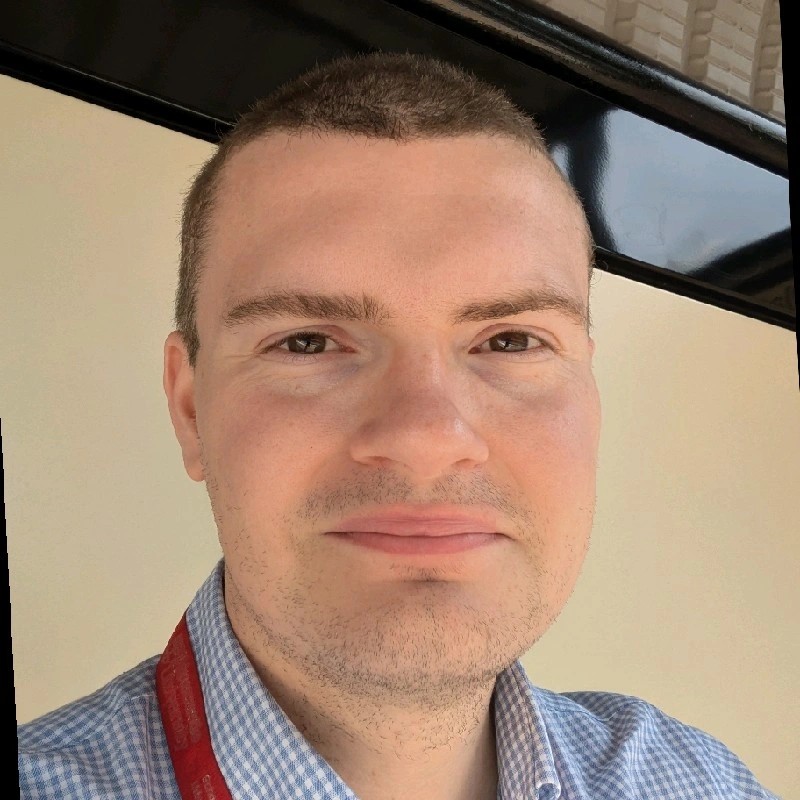}
\textbf{Carlton Shepherd} (B.Sc., Ph.D.) received his Ph.D. Information Security from the Information Security Group at Royal Holloway, University of London, and his B.Sc.\ in Computer Science from Newcastle University. He is currently a Senior Research Fellow at the Information Security Group at Royal Holloway, University of London. His research interests include trusted execution environments and embedded systems security.
\endbio

\end{document}